\documentclass[nofootinbib,twocolumn]{revtex4-1}
\usepackage{booktabs}
\usepackage{xcolor}
\usepackage{graphicx}
\usepackage{graphics}
\usepackage{amsmath}
\usepackage{amssymb}
\usepackage{amsthm,amstext}
\usepackage{hyperref}
\usepackage{amsfonts}
\usepackage{lipsum}
\usepackage{MnSymbol}
\usepackage{mathrsfs}
\usepackage{stmaryrd}
\usepackage{latexsym}
\usepackage[applemac]{inputenc}
\usepackage{accents}

\newcommand\beq{\begin{equation}}
\newcommand\beqn{\begin{eqnarray}}
\newcommand\eeq{\end{equation}}
\newcommand\eeqn{\end{eqnarray}}

\begin{document}

\title{Strain incompatibility as a source of residual stress in welding \\ and additive manufacturing}       

\author{Domenico Zaza$^a$, Michele Ciavarella$^a$, Giuseppe Zurlo$^b$}

\address{$^a$Dipartimento di Meccanica, Matematica e Management (DMMM), Politecnico di Bari, Via Orabona 4 - 70125 Bari, Italy; $^b$School of Mathematics, Statistics and Applied Mathematics, NUI Galway, University Road, Galway, Ireland.}

\date{October 15, 2020}

\begin{abstract}
The accumulation of residual stress during welding and additive manufacturing is an important effect that can significantly anticipate the workpiece failure. In this work we exploit the physical and analytical transparency of
a 1.5D model to show that the deposition of thermally expanded material onto an elastic substrate leads to the accumulation of strain incompatibility. This field, which is the source of residual stress in the system, introduces memory of the construction history even in the absence of plastic deformations. The model is then applied to describe the onset and the progression of residual stresses during deposition, their evolution upon cooling, and the fundamental role played by the velocity of the moving heat source. 
\end{abstract}

\maketitle

\section{Introduction}

Many natural or technological processes where mass is continuously deposited on the boundary surface of a growing body exhibit a peculiar accumulation of residual stress, like in the growth of plants \cite{Archer,Dumais2001}, in cell motility \cite{Dafalias,johnpre14}, in civil engineering \cite{GoodmanSlopes,Labuz}, but also in planetary formation  \cite{Kadish2005} and in crystallisation \cite{Wildeman,Fink,Schwerdtfeger}. 

\begin{figure}
\centering
\includegraphics[width=\columnwidth]{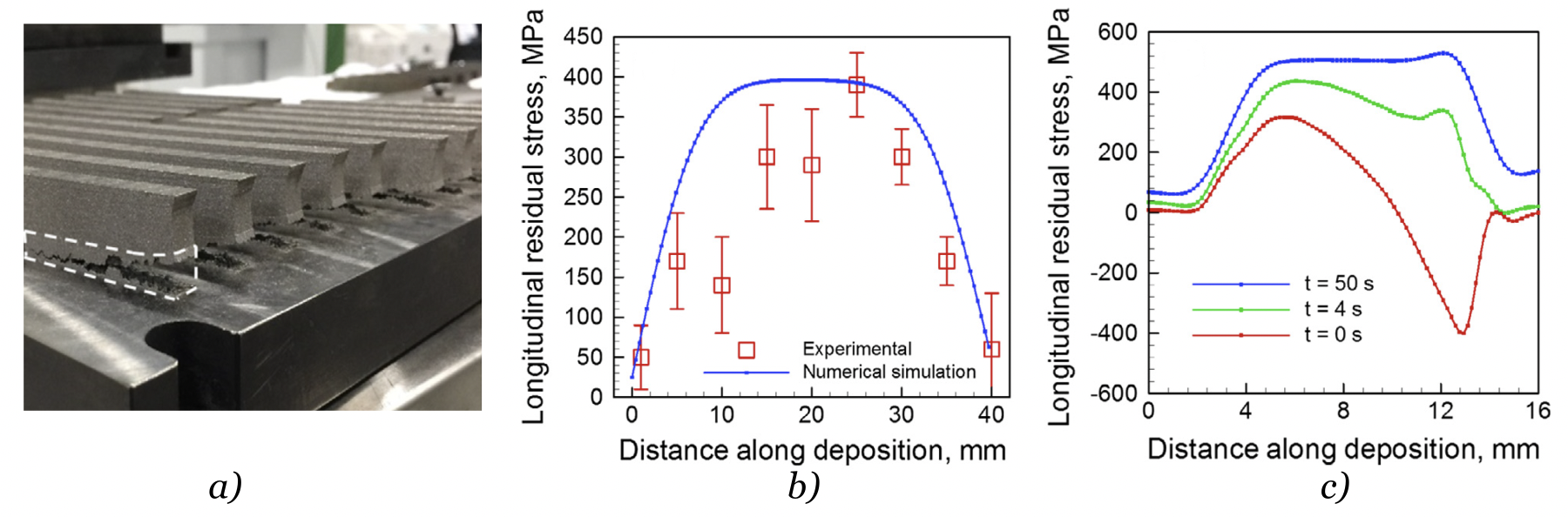}
\caption{$a)$ Failure of an additively manufactured part due to residual stress build-up, courtesy of \cite{To}; $b)$ experimental (whiskers) and computational (solid curve) modelling of longitudinal stress accumulation during additive manufacturing and $c)$ numerical modelling of the evolution of the longitudinal residual stress upon cooling, courtesy of \cite{Mukherjee_2016}.}
\label{intro}
\end{figure}

Welding and additive manufacturing fall in the general category of surface growth problems, since both technologies involve the surface deposition of thermally expanded material and result into the accumulation of longitudinal residual stress at the end of the process \cite{Sames_2016, Mercelis_2006}. Experimental evidence suggests that the magnitude of residual stress generally increases upon cooling, potentially leading to the formation of cracks and to delamination, see Fig.\ref{intro}. Experiments also show that the velocity of the moving heat source plays an important role in the maximum residual stress at the end of deposition \cite{Yadroitsev_2012, Sudersanan2012, Mitra2019,Ravisankar2014,Elmesalamy2016, Liu2015, Apostol2012, Thorat2013, SelectiveLaserManuf, Wang2016}. 

The source of residual stress in welding and additive manufacturing is ``strain incompatibility'': this field measures the local mismatch between stress-free configurations of adjacent layers and is the sole source of residual stress when the workpiece reattains a uniform temperature \cite{Boley, Ueda}. In this work we provide a model to describe how strain incompatibility is accumulated during the deposition of thermally expanded material, even in the absence of other inelastic effects (like plastic deformations or phase transformations). The resulting distribution of strain incompatibility depends on the whole thermal history of the workpiece, thus nailing the memory of the whole manufacturing process in the final distribution of residual stress \cite{Vanel}.

Existing models for the accumulation of residual stress in welding and additive manufacturing are broadly plasticity-based:  in this perspective, temperature gradients induce thermal stresses, which in turn lead to plastic deformations upon attainment of the material yield strength \cite{Leggatt 2008}; the resulting plastic deformations, if incompatible or inherent \cite{Ueda}, are a source of residual stress \cite{Chen1992}. This mechanism was extensively exploited in the numerical modelling of residual stress in additively manufactured stainless steel and alloys \cite{ChenNature,To,Mukherjee_2016,Fergani2016,Buchbinder2014}, in laser-aided additive manufacturing \cite{Ren2018}, and in welding \cite{Sudersanan2012,Ravisankar2014,Liu2015,Apostol2012,Wang2016,Mendes2017}. 

While certainly plasticity plays an important role (above all during solidification, when the stress increases significantly), there is a growing experimental evidence that immediately at the end of deposition, the residual stress can be significantly lower than the limit yield strength of the material, both for welding and for additive manufacturing \cite{Mitra2019,Mukherjee_2016,Lai2013,Wang2015}. Henceforth, it remains to explain, in the absence of plastic deformations, where does this initial distribution of residual stress originates from. 

This perspective is further supported by the fact that in several problems of surface accretion, like in sand-pile growth \cite{Vanel} or in gravitational accretion \cite{GoodmanSlopes,Goodman}, the final body has a distribution of residual stress that can not be explained in terms of plastic yielding, but rather in terms of strain incompatibility stored during the process \cite{ZT1,ZT2,ZT3}, when the deposited material is fully in the elastic range. 

To illustrate this idea in a minimal and mechanically transparent setting, in this work we exploit the simplicity of a toy ``1.5D model''. After identifying the mechanical source of residual stress in strain incompatibility, we describe how this field can be permanently stored during the deposition of thermally expanded material filaments. Strain incompatibility, which quantifies the strain mismatch between the deposited filaments and the underlying substrate, is ``frozen'' at the instant of deposition and does not evolve afterwards. Its value at a given point depends globally on the temperature profile on the bar, and for this reason strain incompatibility is a path-dependent field that depends, in particular, on the velocity of the moving heat source. Our model explains the peculiar shape of the longitudinal residual stress profile at the end of deposition, its evolution upon cooling, and the fundamental role played by the velocity of the moving heat source on the maximum value of the residual stress. 

The manuscript is organised as follows. Strain incompatibility is firstly introduced in a purely mechanical setting in Sec.\ref{SecInc}. We then describe the deposition process in Sec.\ref{Model}, and we show how strain incompatibility can be  determined analytically in the simple case where the elastic moduli are constant. We then address the case of temperature-dependent moduli, that can be approached numerically. In Sec.\ref{SecAppl}, after having introduced an approximated solution to the heat diffusion problem, we apply the model both in the simple case of constant moduli, and in the more realistic case of variable moduli, by highlighting the qualitative and quantitative overlaps with the existing experimental/computational estimates.

\section{Strain incompatibility in a ``1.5D model''\label{SecInc}}

\begin{figure}
\centering
\includegraphics[width=\columnwidth]{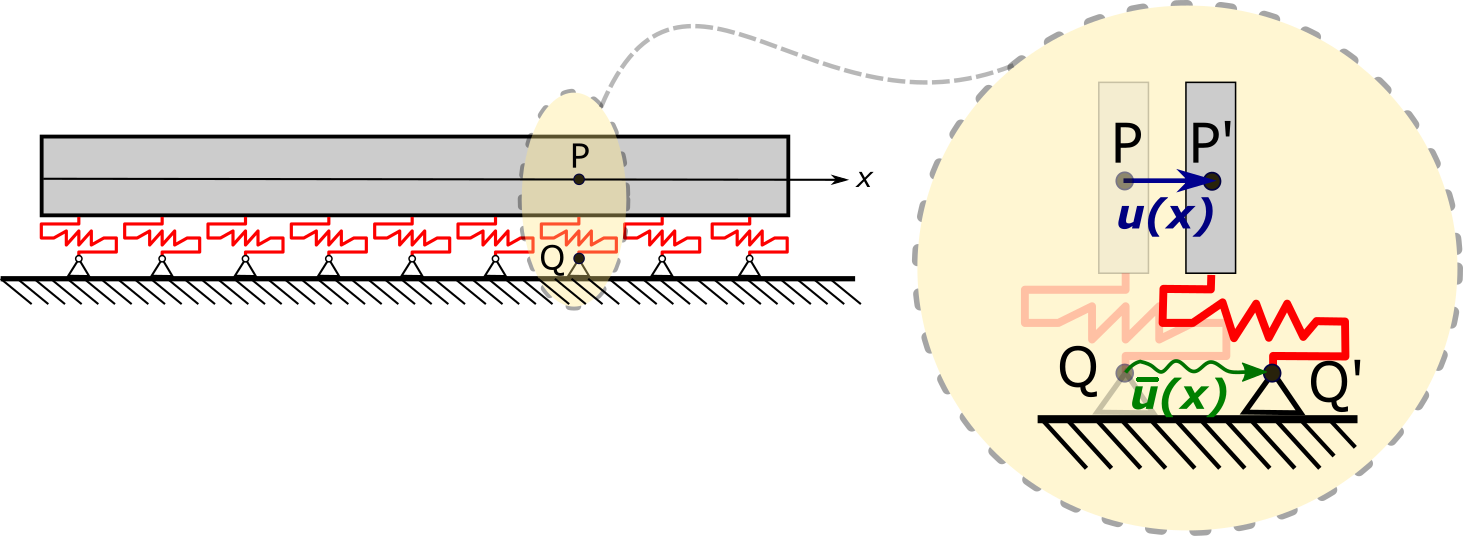}
\caption{Schematic representation of the 1.5D model, made of a 1D elastic rod, that is laterally constrained to an elastic foundation through shear-springs. Here $u(x)=P'-P$ denotes the elastic displacement of the bar, whereas $\bar{u}(x)=Q'-Q$ denotes the imposed displacement of the anchoring point of shear springs.}
\label{mechscheme}
\end{figure}

The minimal setting to describe the presence of residual stress is provided by a ``1.5D model'', describing an elastic 1D bar, laterally constrained by shear springs to a rigid foundation, see Fig.\ref{mechscheme}. In the absence of external loading, by acting on the anchoring points of the shear springs on the rigid foundation it is possible to build up ``strain incompatibility'', which is the sole source of residual stress in the system. After introducing the concept of strain incompatibility in a purely mechanical 1.5D model, we will later show its genesis in the process of deposition, solidification and adhesion to an elastic substrate. 

Consider an elastic rod of length $L$, which is connected to a rigid foundation through shear springs, as represented in Fig.\ref{mechscheme}. Points in the bar are denoted by their referential coordinate $x\in(0,L)$ and their displacement is denoted by $u(x)$, so the strain is $\epsilon(x)=u'(x)$ (where $'$ denotes differentiation with respect to $x$) and the stress is $\sigma(x)=E(\epsilon(x)-\epsilon_i(x))$, with $E$ the bar longitudinal elastic modulus. Here, and in very general terms,  $\epsilon_i(x)$ is a prescribed inelastic strain that could describe, for example, thermal expansion, plastic deformations or phase transformation strains \cite{Ferro1,Ferro2}.

The shear springs exert a force per  unit length $q(x)=G(\bar{u}(x) - u(x))$ on the bar, where $G$ may be given the meaning of a shear modulus and where the field $\bar{u}$ describes the placement of the anchoring point of the shear springs on the rigid foundation, as illustrated in the inset of Fig.\ref{mechscheme}. As well as $\epsilon_i$, assume that such field can be externally controlled. Equilibrium of the bar under null end tractions results into the problem
\beq\label{equi1}
\left\{
\begin{array}{lll}
\sigma'(x) + q(x) = 0 \\
\sigma(x)=E(\epsilon(x)-\epsilon_i(x))\\
\epsilon(x) = u'(x) \\
q(x)=G(\bar{u}(x) - u(x)) \\
\sigma(0) = \sigma(L) = 0. 
\end{array}
\right.
\eeq
To further clarify the role of the inelastic strains $(\epsilon_i, \bar{u}')$ in the accumulation of residual stress, it is convenient to eliminate $u(x)$ in \eqref{equi1} so to obtain a stress-based formulation of equilibrium, 
\beq\label{resstress}
\left\{
\begin{array}{lll}
\displaystyle\sigma''(x)-(G/E)\sigma(x)=G(\epsilon_i(x) - \bar{u}'(x))=:G\eta(x)\\
\sigma(0) = \sigma(L ) = 0. 
\end{array}
\right.
\eeq
In the expression above we have introduced the {\it strain incompatibility} $\eta(x) = \epsilon_i(x) - \bar{u}'(x)$, that is the sole source of residual stress in the bar. In fact, the problem can be solved for a generic distribution of $\eta$ and we obtain
\beqn
&&\sigma=\frac{\beta E}{2}\times\label{analytsolz}\\
&&\left\{[\coth (\beta  L)-1] \sinh (\beta  x) \int_0^L \eta\, e^{\beta  s}  \, ds -e^{-\beta  x} \int_0^x \eta\, e^{\beta  s}  \, ds + \right.\nonumber\\
&& \left. -[\coth (\beta  L)+1] \sinh (\beta  x) \int_0^L \eta\, e^{-\beta  s}  \, ds+e^{\beta  x} \int_0^x \eta \,e^{-\beta  s}  \, ds \right \}.\nonumber
\eeqn
If $\eta=0$ there is no stress in the system, whereas if $\eta\neq 0$, the resulting distribution of $\sigma$ will be nontrivial. The field $\eta$ here introduced may be seen as the 1.5D counterpart of the 3D tensorial strain incompatibility which is, also in that general context, the sole source of residual stress in the absence of external loading \cite{ZT1,ZT2,ZT3}. We note that the term {\it incompatibility} stems from the fact that $\eta=0$ is implied by $\epsilon_i=\bar{u}'$, meaning that a non-trivial distribution of $(\bar{u}', \epsilon_i)$, if {\it compatible}, does not produce stress in the system \cite{Boley}. 

With reference to welding and additive manufacturing, while $\epsilon_i$ may be given the meaning of a thermal expansion during the deposition of hot material, the field $\bar{u}'$ traduces the shear mismatch between the bar and the underlying layer. As we show in the next section, while $\epsilon_i$ vanishes when the temperature of the bar is uniform and equal to the ambient temperature, the field $\bar{u}$ is ``frozen'' at the instant of deposition and it does not evolve afterward. Such field will persist at the end of manufacturing, producing residual stress without necessarily involving the onset of plastic deformations. 

We finally remark that the 1.5D model directly introduced above is not derived from a parent 3D theory. For this reason, our model aims only at a qualitative description of the fundamental effects taking place during the deposition and solidification.

\section{Genesis of incompatibility during deposition\label{Model}}

We now describe the process of accretion of a 1.5D bar through lateral deposition of mass on its ``growth end''. We introduce an evolving {\it reference configuration} $\mathcal{R}_t=\{x\in(0,\psi(t))\}$, where $t$ represents a time-like parameter and where $\psi(t)$ denotes the evolution of the referential growth end. We further denote by $D_0(t)=\dot\psi(t)$ the referential velocity of the growth end, where the dot denotes differentiation with respect to $t$.  By assuming that the referential mass density $\varrho_0$ is uniform, the rate of deposited mass is $\mu=D_0\varrho_0$. Since $\psi(t)$
is an increasing function, we can define its inverse $\vartheta=\psi^{-1}$ so that $t = \vartheta(x)$ gives the instant when a material point $x$ is located on the growth end.

To simplify the analysis we assume that heat diffusion is uncoupled from the equilibrium, so we assume that the temperature profile $T(x,t)$ in the bar is always known during the process. In the subsequent applications of the theory, we will adopt  temperature profiles typically suggested in literature to model the process of welding and additive manufacturing. 

During the deposition of hot material, a referential fibre $dx$ will be deposited in a thermally expanded length $dy=dx(1+\epsilon_i(x,t))$, with $\epsilon_i(x,t) = \alpha(T(x,t)) \Delta T(x,t)$, where $\alpha$ is the coefficient of thermal expansion and $\Delta T(x,t)=T(x,t) - T_{amb}$, where $T_{amb}$ is a reference ambient temperature. Due to thermal expansion and to the build-up of elastic stress, points of the bar can undergo a time-dependent displacement $u(x,t)$, which is assumed small together with the strain $\epsilon(x,t)=u'(x,t)$. The current placement of the growth end is given by $\xi(t)=\psi(t)+u(\psi(t),t)$, so that the actual velocity of the growth end is $D(t) = \dot{\xi}(t) = D_0(t) + \dot{u}(\psi(t),t) + D_0(t) \epsilon(\psi(t),t)$. By assuming that $\epsilon\ll 1$ and that $\dot{u}\ll D_0$, the current and referential velocities $D$ and $D_0$ may be confused for all practical purposes. We finally introduce the displacement of the growth end at the deposition instant,
\beq
\bar{u}(\psi(t)) = u(\psi(t),t) = \xi(t) - \psi(t), 
\eeq
which is a main player in our model. Note that by making use of the function $\vartheta(x)$, it results $\bar{u}(x)=u(x,\vartheta(x)) = \xi(\vartheta(x)) - \psi(\vartheta(x))$. 

\begin{figure}
\centering
\includegraphics[width=\columnwidth]{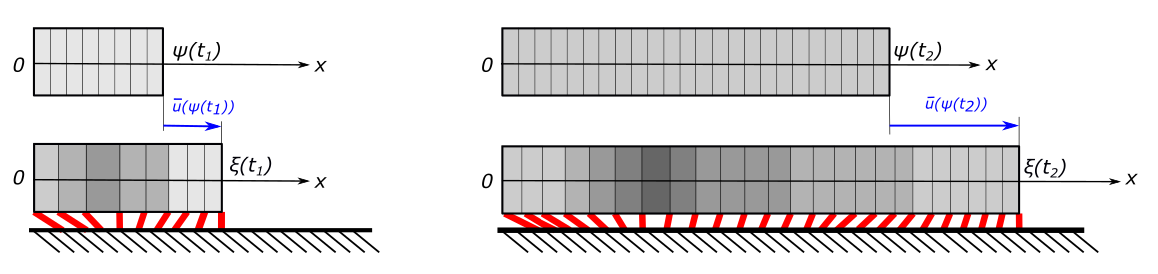}
\caption{Schematic representation of the reference (above) and current (below) configurations of the bar in two instants $t_1$ and $t_2>t_1$. The springs connecting the bar to the foundation, denoted in red in the current configurations, are un-sheared at deposition. The scheme also illustrates the role of the field $\bar{u}$ in defining the current placement of the growth-end.}
\label{reference}
\end{figure}

To impose equilibrium at each stage of accretion, further assumptions are required on the mechanical state of the system. The first assumption is that the longitudinal elastic modulus $E$ and the shear modulus $G$ are known functions of temperature; concerning their ratio, since in 3D elasticity it results that $G/E=1/(2(1+\nu))$ where $\nu$ is the Poisson coefficient, and since for a wide class of materials $\nu$ is essentially independent on temperature \cite{Liu2015,Wang2016,Farid 2013}, we here assume that $\beta^2=G/E$ is constant. We assume, as well, that the thermal expansion coefficient $\alpha$ does not depend on temperature \cite{Liu2015,Wang2016,Farid 2013}.

To describe the deposition and contextual adhesion to the substrate, we assume that new ``drops'' of material are continuously deposited in a close-to-melted state, in a traction-free state ($\sigma(\psi(t),t)=0$). As soon as temperature decreases below the deposition temperature $T_{dep}$, we assume that the material instantaneously adheres to the foundation. This means that the anchoring point of shear springs on the rigid foundation coincides with the current placement of the growth end $\xi(t)=\psi(t)+\bar{u}(\psi(t))$, so that the resulting shear force is $q(x,t) = G(T(x,t))(\bar{u}(x) - u(x,t))$. Due to the definition of $\bar{u}(x)$, this means that the shear force between the substrate and the bar is zero at deposition ($q(\psi(t),t)=0$). 

To sum up, equilibrium of the growing bar can be imposed, by slight modification of \eqref{equi1}, as
\beq\label{equi2}
\left\{
\begin{array}{lll}
\sigma'(x,t) + q(x,t) = 0 \\
\sigma(x,t)=E(T(x,t))(\epsilon(x,t)-\alpha(T(x,t))\Delta T(x,t)) \\
\epsilon(x,t) = u'(x,t)\\
q(x,t)=G(T(x,t)) (\bar{u}(x) - u(x,t)) \\
\sigma(\psi(t),t) = 0\\
\sigma(0,t)=0. 
\end{array}
\right.
\eeq
The structure of this problem is similar to the purely mechanical theory of 1.5D surface growth proposed in \cite{ZT2}. Since temperature is prescribed during the process, the main unknown in the problem is represented by the function $\bar{u}'(x)$ which, as described by \eqref{resstress}, is the sole source of residual stress at the end of deposition, when the bar is at ambient temperature.  

To solve this problem we compute the field $u(x,t)$ as a function of the yet unknown field $\bar{u}(x)$, and we thus impose the consistency condition $\bar{u}(x)=u(x,\vartheta(x))$, leading to an integral equation for $\bar{u}'(x)$, see \cite{ZT2} for a similar analysis. An analytically simpler, and physically more transparent, incremental formulation of this problem is discussed in the next section.

We remark that the assumption that strain incompatibility $\bar{u}'$ is ``frozen'' at the instant of deposition and does not evolve afterwards may always be relaxed to permit, for example, the description of delamination effects or the onset of plastic deformations during deposition. The analysis of these effects are left for future studies.

\subsection{Incremental formulation: constant moduli\label{SecIncr}}

The general problem discussed above can be conveniently recast in an incremental form, that we first discuss under the simplifying assumption $E, G$ and $\alpha$ are constant. In this case, the problem admits a fully analytical solution, that provides a good qualitative description of the main effects observed in literature. We later focus on the case where the elastic moduli depend on temperature. 

 An incremental formulation of problems involving surface growth was originally proposed by \cite{Goodman,BG1} in the context of gravitational accretion, and by \cite{KF} in the context of water freezing. Later developments based on the seminal work of \cite{Trincher} paved the way to general formulations of the theory of surface growth, both in the linear \cite{ZT1} and in the nonlinear \cite{ZT3} setting. The current formulation is inspired by \cite{ZT2}, where a purely mechanical theory of a growing 1.5D bar was presented. 

The incremental formulation of \eqref{equi2} relies upon on the possibility to recast the displacement and the stress in the bar, at an arbitrary instant $t\geq \vartheta(x)$ after deposition, as
\beq \label{sigmaincr}
\sigma(x,t) = \sigma(x,\vartheta(x)) + \int_{\vartheta(x)}^t\dot\sigma(x,s)\,ds
\eeq
\beq \label{uincr}
u(x,t) = u(x,\vartheta(x)) + \int_{\vartheta(x)}^t\dot{u}(x,s)\,ds 
\eeq
where we recall that $\sigma(x,\vartheta(x))=0$ and $u(x,\vartheta(x))\equiv\bar{u}(x)$. Thus, the equilibrium equation \eqref{equi2}$_1$ can be recast as
\beq\label{interm}
 - \dot{\sigma}(x,\vartheta(x))\vartheta'(x) + \int_{\vartheta(x)}^t\left(\dot\sigma'(x,s) - G\,\dot{u}(x,s)\right)\,ds = 0. 
\eeq
For $x=\psi(t)$, the equation above delivers an incremental boundary condition $\dot\sigma(\psi(t),t)=0$. Since \eqref{interm} holds for all $t\geq\vartheta(x)$, by localisation we obtain the incremental equilibrium equation $\dot\sigma'(x,t) - G\,\dot{u}(x,t) = 0$. Together with time differentiation of \eqref{equi2}$_{2,3,6}$ we thus obtain the incremental formulation of the equilibrium problem  \eqref{equi2}, 
\beq\label{equi4}
\left\{
\begin{array}{lll}
\dot\sigma'(x,t) -G\,\dot{u}(x,t) = 0\\
\dot\sigma(x,t)=E(\dot\epsilon(x,t)-\alpha\dot{T}(x,t))\\
\dot\epsilon(x,t) = \dot{u}'(x,t)\\
\dot\sigma(\psi(t),t) = 0\\
\dot\sigma(0,t)=0. 
\end{array}
\right.
\eeq
The advantage of \eqref{equi4} with respect to \eqref{equi2} is that the incremental displacement $\dot{u}(x,t)$ does not depend on the yet unknown field $\bar{u}'(x)$, whereas $\bar{u}'(x)$ appears explicitly in the problem \eqref{equi2}. Furthermore, at least in the special case when the elastic moduli are constant, the incremental solution $\dot{u}$ can be determined analytically. 

Once the field $\dot{u}$ is known, the resulting distribution of $\bar{u}'(x)$ can be obtained by recalling that $\bar{u}'(x)=\epsilon(x,\vartheta(x))+\vartheta'(x)\dot{u}(x,\vartheta(x))$. By using this identity together with the condition $\sigma(x,\vartheta(x))= 0$, we finally obtain
\beq\label{ducalc}
\bar{u}'(x) = \underbrace{\alpha(T_{dep}-T_{amb})}_{\text{local countepart}} \,\, + \,\,\underbrace{\vartheta'(x)\dot{u}(x,\vartheta(x))}_{\text{non-local countepart}}. 
\eeq
This expression shows that the incompatibility resulting from deposition and adhesion to an elastic substrate is made of two counterparts: a {\it local}, uniform contribution, that is controlled by the deposition temperature; and a {\it non-local} contribution, that depends on the incremental displacement of the growth-end and, henceforth, on the deformation of the whole bar. Due to the non-local term, deposition protocols that share the same deposition temperature, but that differ in terms of spatial distribution of temperature, will ultimately result into different distributions of residual stress. In this sense, the strain incompatibility permanently accumulated through $\bar{u}'$  holds the memory of the construction history \cite{Vanel}, a path-dependence effect already highlighted in the context of purely mechanical surface growth  \cite{ZT1,ZT2,ZT3}. 

Under the assumption of constant velocity, $\vartheta(x) = V x$, the solution $\dot{u}$ to the incremental problem can be found analytically and then, through \eqref{ducalc}, we finally obtain, for a generic distribution of temperature in the bar, 
\beqn
&&\bar{u}'(x)=\alpha\times\nonumber\\
&&\left[ T_{dep}-T_{amb} + \, \frac{\text{csch}(\beta  x)}{V} \int_0^x \dot{T}(z,\vartheta(x)) \sinh (\beta  z) \, dz \right].\label{Analytical1}\hspace{12pt} 
\eeqn
This expression, which is our main finding, shows explicitly how the field $\bar{u}'(x)$ depends on the whole previous thermal history $T(z\leq x,s\leq \vartheta(x))$ in the bar. The resulting distribution of $\bar{u}'(x)$ can be used to compute the evolution of stress during deposition, by using the analytical expression \eqref{analytsolz} together with $\eta(x,t) = \alpha\Delta T(x,t) - \bar{u}'(x)$. To compute the residual stress in the bar at end of accretion, when the bar is uniformly at the ambient temperature, we can use  \eqref{analytsolz} with $\eta(x) =  - \bar{u}'(x)$.

\subsection{Incremental formulation: temperature dependent moduli\label{SecIncr2}}

The assumption of constant moduli discussed above leads to an underestimate of the stress accumulation during the deposition. For this reason, we now briefly discuss how the incremental approach discussed above can be modified to account for temperature dependence of the moduli.  In this case, the incremental equilibrium equation \eqref{equi2}$_{1,4}$ rewrites as
\beq
\dot{\sigma}'(x,t)- G(T(x,t)) \dot{u}(x,t)= \dot{G}(T(x,t)) (u(x,t)-\bar{u}(x)). 
\eeq
Likewise, the incremental form of the constitutive equation \eqref{equi2}$_{2}$ is
\beqn
&&\dot{\sigma}(x,t) = E(T(x,t)) (\dot\epsilon(x,t) - \alpha  \dot{T}(x,t) )\nonumber \\
&& \hspace{40pt}+ \dot{E}(T(x,t))  ( \epsilon(x,t) -\alpha  \Delta T(x,t)).  
\eeqn
The resulting incremental formulation of the equilibrium problem with temperature dependent moduli is 
\beq\label{equi5}
\left\{
\begin{array}{lll}
\dot{\sigma}'(x,t)- G(T(x,t)) \dot{u}(x,t)= J(x,t)\\
\dot{\sigma}(x,t) = E(T(x,t)) ( \dot\epsilon(x,t) - \alpha  \dot{T}(x,t) ) + K(x,t)\\
\dot\epsilon(x,t) = \dot{u}'(x,t)\\
\dot\sigma(\psi(t),t) = 0\\
\dot\sigma(0,t)=0
\end{array}
\right.
\eeq
where the functions 
\beqn
&&J(x,t)=\dot{G}(T(x,t)) \int_{\theta(x)}^{t} \dot{u}(x,s)ds,\nonumber\\ 
&&K(x,t)=\frac{\dot{E}(T(x,t))}{E(T(x,t))}  \int_{\theta(x)}^{t} \dot{\sigma}(x,s)ds\nonumber
\eeqn
depend on the previous history of deformation. In this case the incremental solution $\dot{u}$ cannot be determined analytically and the problem can be approached by a finite difference formulation. The resulting incompatibility can be computed, once again, by using \eqref{ducalc}.

\section{Applications\label{SecAppl}}

The model can now be applied to reproduce some important features observed in numerical and experimental studies on welding and additive manufacturing. We separately present the fully analytical results in the case of constant moduli, and the numerical results relative to non-constant moduli. 

\subsection{Temperature distribution during deposition}

\begin{figure}
\centering
\includegraphics[scale=0.5]{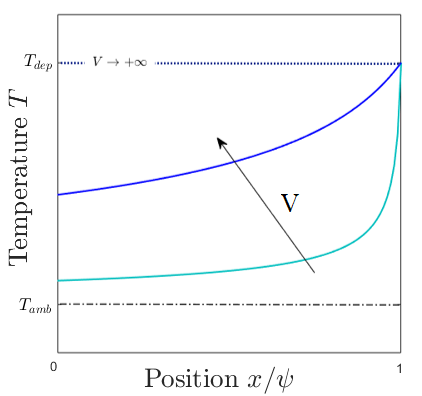}
\caption{The dependence of temperature in the workpiece on the heat source velocity $V$, as predicted by the modified Rosenthal solution \eqref{Ros}.}.
\label{rosenthal}
\end{figure}

To account for the velocity of the moving heat source we embrace the approximated Rosenthal solution to the heat diffusion problem \cite{Rosenthal}, frequently adopted in the modeling of welding and additive manufacturing \cite{rosenthalam}. Such solution is here modified to avoid temperature singularities at deposition. By assuming that the velocity $V$ of the moving heat source is constant and equal to the deposition velocity $D_0$, so that $\psi(t)=V t$, we take 
\beq\label{Ros}
T(x,t) = T_{amb}+a\sqrt{\frac{V}{\psi(t)-x+w}}
\eeq
where the tuning parameter $a$ is proportional to the specific heat flux of the source, and where $w=a^2 V/(T_{dep}-T_{amb})^2$. The resulting dependence of temperature on velocity, qualitatively reported in Fig.\ref{rosenthal}, shows that higher velocities of the moving heat source result into higher temperature of the whole bar. This confirms the physical expectation that if the heat source moves faster, there is less time for heat transfer to take place with the surrounding material. 

\subsection{Predictions in the case of constant moduli}

While in realistic applications of welding and additive manufacturing the elastic moduli have a strong variation with temperature, we can still show that under the simplifying assumption of constant elastic moduli our theory gives good qualitative agreement with the experimental observations. The advantage is, in this case, that the solution is fully analytical. At this point, by adopting the temperature profile \eqref{Ros}, we can introduce the dimensionless stress in the bar
\beq\label{dimlesssigma}
\tilde\sigma = \sigma / E_{amb}
\eeq
where $E=E_{amb}$ is the homogeneous reference longitudinal modulus. Note that $\tilde\sigma$ depends on the list of parameters $\beta,\alpha,a,(T_{dep}-T_{amb}), V$. 
\begin{figure}
\includegraphics[scale=0.33]{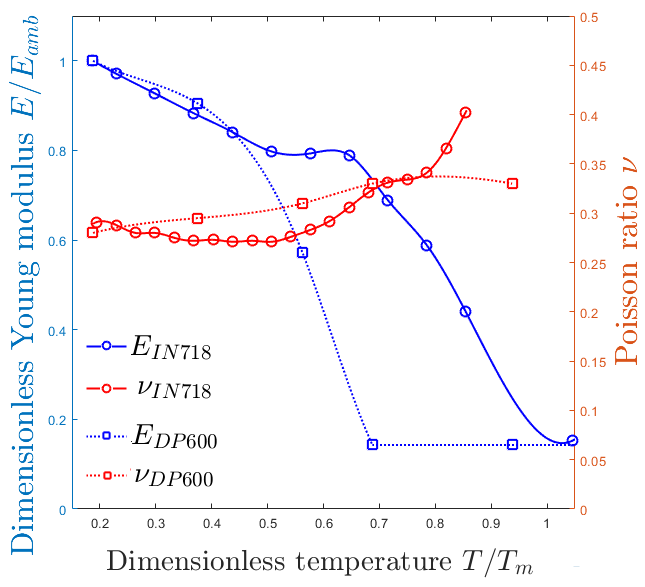}
\caption{Trends of the elastic moduli with temperature for two materials: IN718 (used in AM) data from \cite{Mukherjee_2016} and DP600 (used in welding) data from \cite{DP600prop1,DP600prop2}}.
\label{elastictrendv}
\end{figure}

\begin{figure*}
\includegraphics[scale=0.5]{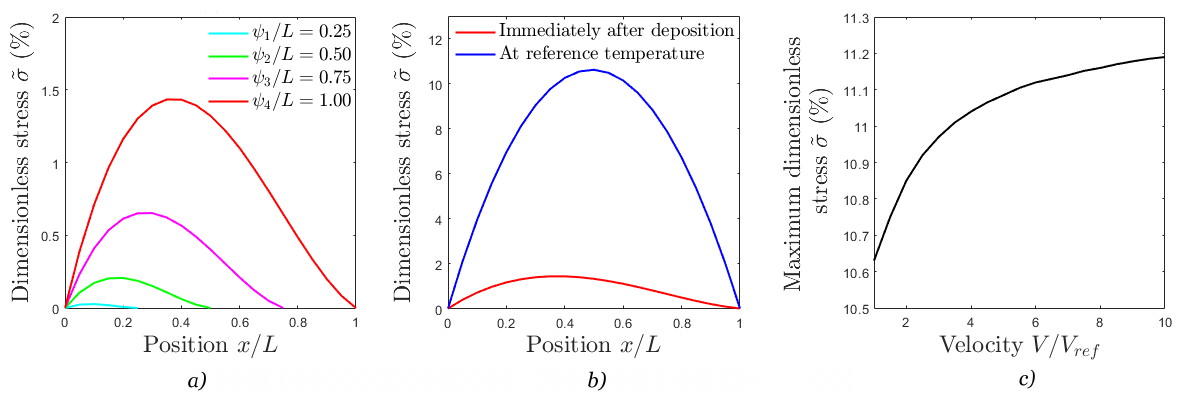}
\caption{Analytical results of the 1.5D model using  the moving temperature profile \eqref{Ros}: (a) dimensionless stress distributions in four intermediate stages of deposition for a fixed value of velocity $V=1$ using \eqref{analytsolz} together with  $\eta = \alpha\Delta T - \bar{u}'$; (b) dimensionless stress distributions immediately after deposition and at reference temperature for a fixed value of velocity $V=1$, obtained by using \eqref{analytsolz} together with  $\eta= - \bar{u}'$; (c) maximum dimensionless stress for growing velocity $V$ of the moving heat source. We have used: $L=1$, $T_{dep}-T_{amb}=1$, $E_{amb}=1$, $\alpha=1$, $a=1$, and $\beta=1$.}
\label{Analytical}
\end{figure*}

The resulting stress profile immediately at the end of deposition, its modification during cooling and the influence of velocity on the maximum residual stress are given by the fully analytical expressions \eqref{analytsolz},\eqref{Analytical1} together with the temperature profile \eqref{Ros}. The results, represented in Fig.\ref{Analytical}, show that during deposition the stress is non-symmetric with respect to the bar midpoint, and that the stress increases as deposition advances, see Fig.\ref{Analytical}$_a$. At the end of deposition, the temperature decreases while the strain incompatibility is frozen at its deposition value \eqref{Analytical1}. As a result, the stress in the bar increases significantly upon cooling, as illustrated in Fig.\ref{Analytical}$_b$. Finally, the maximum residual stress in the bar is an increasing function of the velocity of the moving heat source, see Fig.\ref{Analytical}$_c$. As discussed later on, all these trends are recovered in the case of variable elastic moduli, and are in reasonable qualitative agreement with the available experimental and computational observations. 

\subsection{Predictions in the case of temperature dependent moduli}

In most materials used in welding and additive manufacturing, both the longitudinal and the shear elastic moduli are decreasing functions of temperature, whereas their ratio (which is a function of the Poisson coefficient), is essentially constant over a wide range of temperatures, see Fig.\ref{elastictrendv} \cite{Liu2015,Farid 2013, Wang2016}. To account for the reduction with temperature we here assume that $E(T)=E_{amb}(1-r(T-T_{amb}))$, where $E_{amb}$ is the value of the longitudinal modulus at the ambient temperature and where $r$ is a tuning parameter. Concerning the shear modulus, we assume that $G(T)=\beta^2 E(T)$ where, in analogy with the 3D theory, $\beta$ is constant. 

The only parameter that can not be extracted from available data at the 3D level is $\beta$: a consistent 3D$\rightarrow$1.5D derivation of $\beta$ is still possible by using methods of  dimension reduction \cite{DPZ2008}, but this task is out of the scopes of our work; for this reason, we here treat it as a tuning parameter. More in detail, after obtaining from literature the value of the material/process parameters $\alpha,a,(T_{dep}-T_{amb}),E_{amb},r,V$, we tune $\beta$ on the maximum value of the dimensionless stress $\tilde\sigma$. We then keep $\beta$ fixed when changing other process parameters, such as velocity or temperature profile. Note that with temperature dependent moduli there are no closed form solutions, so the approach described in Sec.\ref{SecIncr2} must be adopted. 

\begin{figure}
\centering
\includegraphics[scale=0.34]{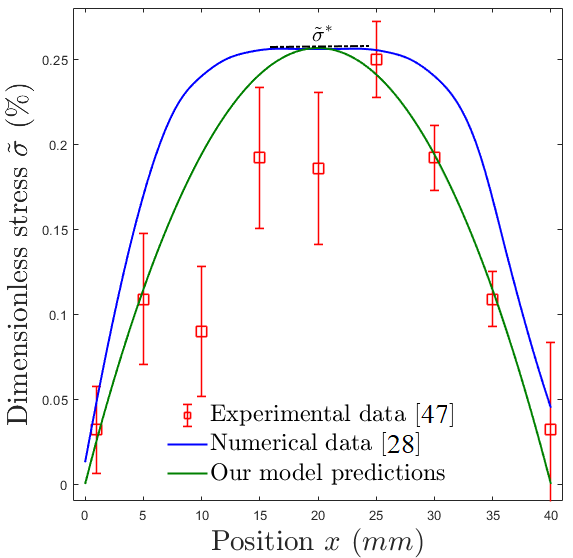}
\caption{\label{exp_res} Residual stress resulting from additive manufacturing of a single layer of alloy on a substrate. The marks report the experimental measurements of \cite{Shah 2014}, from which we have extracted $V=4$mm/s, $L=40$mm, $T_{amb}=300$K, $T_{dep}=1600$K and $a=1056.8 K\sqrt{s}$, $E_{amb}=165$GPa, $r=5.14\cdot 10^{-4}K^{-1}$ and $\alpha=1.43\cdot 10^{-5} K^{-1}$. The blue curve is the numerical estimate of \cite{Mukherjee_2016}, whereas the green curve is the residual stress profile according to our own model, with $\beta=0.0305\,\text{mm}^{-1}$.}
\end{figure}

In our first application, we deal with the stress profile at the end of deposition and we compare our results with the numerical and experimental results of \cite{Mukherjee_2016,Shah 2014}. By extracting the temperature profile from 
 \cite{Mukherjee_2016}, we have tuned the model parameter $\beta$ to achieve the same maximum value of the dimensionless stress $\tilde\sigma$, see Fig.\ref{exp_res}. Our resulting profile for the longitudinal stress is in good qualitative agreement with the experimental measurements, when temperature in the workpiece is uniformly equal to the ambient temperature.  

\begin{figure}
\centering
\includegraphics[width=\columnwidth]{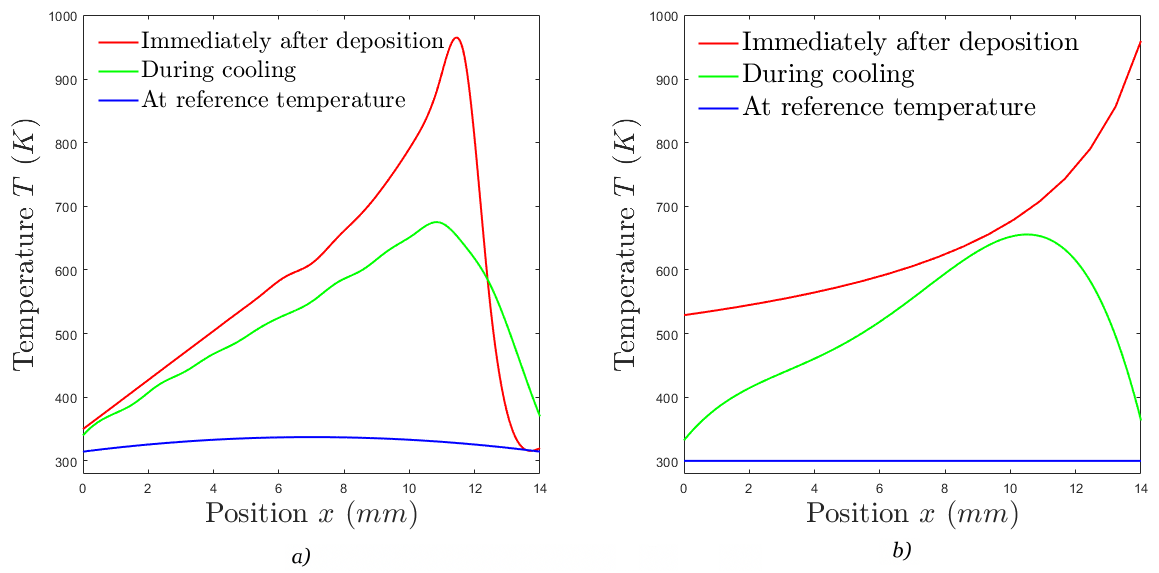}
\caption{(a) Temperature profile, numerically obtained in \cite{Mukherjee_2016} immediately at the end of deposition (red), during cooling (green) and towards the end of cooling (blue). (b) (red) our calibration of the modified Rosenthal solution \eqref{Ros} based on the numerical temperature profile at the end of cooling, for $V=15$mm/s, $L=14$mm, $T_{amb}=300$K, $T_{dep}=960$K, that gives $a=235.9\,\text{K}\text{s}^{1/2}$; (green) polynomial interpolation of the corresponding curve in (a); (blue) the uniform ambient temperature in the bar.}
\label{appl1temp}
\end{figure}

\begin{figure}
\centering
\includegraphics[width=\columnwidth]{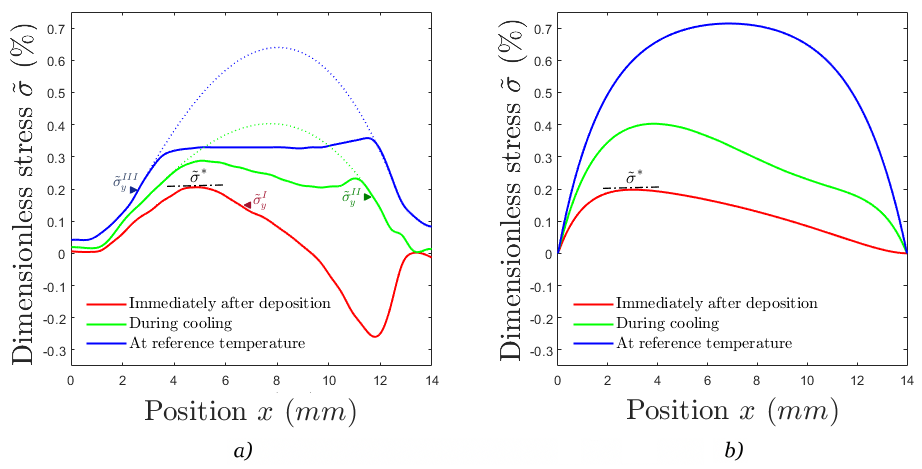}
\caption{(a) Dimensionless stress $\tilde\sigma$, adapted from the numerical results of \cite{Mukherjee_2016} with $E_{amb}=166$GPa. The three stress profiles refer to the three temperature profiles reported in Fig.\ref{appl1temp}$_a$ and are relative to the end of deposition (red),  cooling (green) and ambient temperature (blue). Here $\tilde\sigma_y^{\alpha}$ with $\alpha=${\it I,II,III} are the yield stresses at various temperatures and the dotted curves are $3^{rd}$ order extrapolations of the elastic stress in the plastically deformed zone. (b) Dimensionless stress $\tilde\sigma$ from our model, relative to the temperature profiles reported in Fig.\ref{appl1temp}$_b$. The value $\beta=0.49\text{mm}^{-1}$ was calibrated to achieve the same maximum stress $\tilde\sigma^*$ at the end of deposition in our model and in the numerical simulations.}
\label{appl1stress}
\end{figure}

A second application describes the evolution of the stress profile upon cooling. In this case, we first tune the temperature profile at the end of deposition and during cooling on the numerical solution to the heat diffusion problem given in \cite{Mukherjee_2016} (see Fig.\ref{appl1temp}). We then deduce the distribution of stress in the bar as described in Sec.\ref{SecIncr2}. Upon cooling, the stress in the bar increases and tends to become symmetric, both according to the numerical predictions of \cite{Mukherjee_2016} (see Fig.\ref{appl1stress}$_a$) and according to our model (see Fig.\ref{appl1stress}$_b$). 
Note that numerical predictions of \cite{Mukherjee_2016} show, upon cooling, a flat region due to the attainment of the yield stress, and a clear hardening effect which makes the model quite more elaborate than the present simple, but transparent one. Since we neglect plasticity effects in our model, we cannot expect a quantitative agreement with the numerical predictions of \cite{Mukherjee_2016}, but we would like to point out that the ``origin" of residual stresses, even in the model of \cite{Mukherjee_2016}, remains mostly elastic: as Fig.\ref{appl1stress}$_a$ shows, immediately at the end of deposition (red curve) the stress is mostly below the yield stress for the given temperature. In our perspective, this initial distribution of residual stress is due to the accumulation of strain incompatibility. Then, during cooling, larger plasticity effects appear in the system. 

To attempt a qualitative comparison of our findings with the ones of \cite{Mukherjee_2016}, we have used third order polynomials to extrapolate elastic stress profiles in this work. Based on the values of the yield stresses $\tilde\sigma_y^{\it II}$ and $\tilde\sigma_y^{\it III}$ for the green and the blue curves of Fig.\ref{appl1stress}$_a$, respectively, we have computed the 4 coefficients of the third-order polynomial $\tilde\sigma^{\text{ela}}$ by imposing  that the extrapolated elastic stress profile (dotted curves) has same values and slopes of the numerical profiles in the two points where the yield stress is attained. The comparison of our results with the ones of \cite{Mukherjee_2016}  shows a reasonably good estimate of the stress increments during the cooling process.

\begin{figure}
\includegraphics[width=\columnwidth]{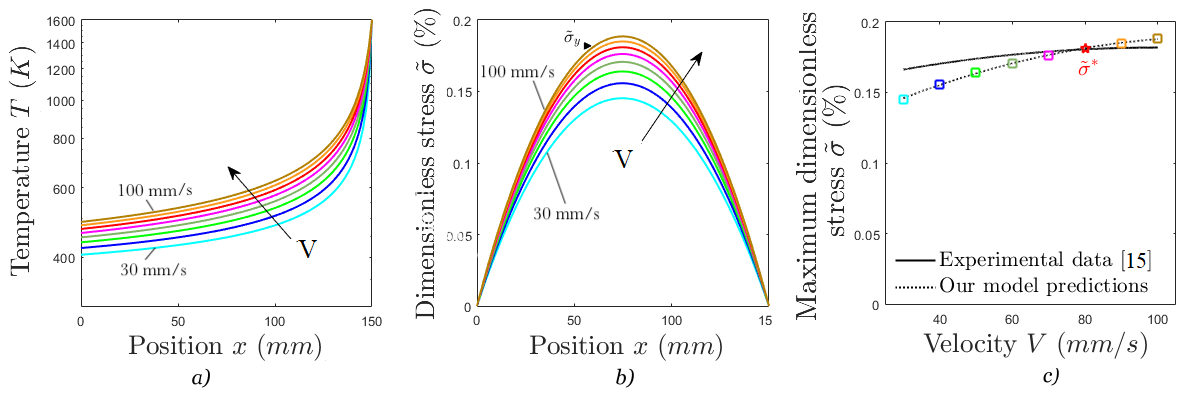}
\caption{\label{stressvelocity} Role of velocity of the heat source on the maximum residual stress, for a constant heat input per unit length. (a)  temperature profiles in the bar for increasing source velocities. (b) Final distribution of residual stress in the bar according to our model, for increasing source velocities. In both figures, the curves range from a velocity of $30$ mm/s (lower curve) to $100$ mm/s (upper curve), with equal steps of 10 mm/s. (c) Experimental (solid) versus our numerical estimates (dotted) for the maximum residual stress versus velocity. Here $L=150$mm, $T_{amb}=300$K, $T_{dep}=1600$K, $a=236.7 K s^{1/2}$, $\alpha=1.82\cdot10^{-5}\text{K}^{-1}$, $E_{amb}=210$GPa, $r=6.124\cdot10^{-4}\text{K}^{-1}$. The value of $\beta=0.0067\,\text{mm}^{-1}$ was tuned to match the marked value of dimensionless stress, relative to $V=80$mm/s.  }
\end{figure}

We finally focus on the effect of velocity on the maximum residual stress. This effect was experimentally studied by \cite{Mitra2019} in the case of laser welding of dual phase steel for a constant heat input per unit length. In this case, experiments show that the maximum residual stress at the end of welding is 
lower than the material yielding stress, supporting our perspective that residual stresses can have a non-plastic origin. 
By extracting from \cite{Mitra2019} the model parameters and by tuning $\beta$ on a reference value of the dimensionless stress, our results, illustrated in Fig.\ref{stressvelocity}, show that higher velocity results into higher maximum stress in the bar: this result is consistent with the experimental findings of  \cite{Mitra2019}, further supporting the capacity of our model to describe a plethora of effects, even in the absence of plastic deformations. 

\section{Conclusions}

In this work we have presented a simple mechanical model that describes the process of deposition and adhesion of thermally expanded filaments to an elastic substrate, and we have applied our model to reproduce some trends observed in numerical and experimental studies on welding and additive manufacturing. Despite of its intrinsic limitations, our 1.5D model correctly predicts several effects observed both in experimental and computational studies, such as the residual stress pattern immediately after deposition, its evolution upon cooling, and the role played by the velocity of the moving heat source on the maximum value of the residual stress. 

Noteworthy, our model does not require the introduction of plastic deformations. While on the one hand plasticity effects can be certainly included, our main task was to show that the inelastic source of residual stress during the deposition of thermally expanded material is represented by strain incompatibility. The relevance of this field was recently discussed for all processes of surface deposition of mass \cite{ZT1,ZT2,ZT3}, above all in the ``inverse engineering'' perspective to embed a required distribution of residual stress through suitable deposition protocols. Future extensions of the idea proposed in this work will require a fully coupled 3D thermo-elastic formulation of the theory, thus paving the way for the possibility to ``design residual stress patterns'' through layered deposition of mass: an exciting perspective for a future generation of 3D printers, capable to embed information not only in the form of elastic properties \cite{Andrei}, but also in the form of prestress \cite{Danescu_2013,Ge}.

\end{document}